\documentclass[12pt]{article} 
\usepackage[english]{babel}
\usepackage{graphicx}
\usepackage{amsmath}
\usepackage{amssymb}
\usepackage{mathtools}
\usepackage{braket}
\usepackage{setspace}
\usepackage[utf8]{inputenc}
\usepackage[makeroom]{cancel}
\usepackage[misc]{ifsym}
\usepackage{float}
\usepackage{placeins}
\usepackage{soul}
\usepackage[normalem]{ulem}

\usepackage[top=2.5cm, bottom=2.5cm, left=2.5cm, right=2.5cm]{geometry}

\def\brack#1{ \left(   #1 \right) }

\def\blockbrack#1{ \left[   #1 \right] }

\numberwithin{equation}{section}

\usepackage{color}
\usepackage[colorlinks=true,
            linkcolor=red,
            urlcolor=blue,
            citecolor=lightblue]{hyperref}
\definecolor{darkblue}{rgb}{0.0, 0.0, 0.5}
\definecolor{darkblue}{RGB}{0,0,80}
\definecolor{lightblue}{RGB}{50,120,150}
\definecolor{darkgreen}{RGB}{0,80,0}
\definecolor{darkred}{RGB}{80,0,0}
\definecolor{amber}{rgb}{1.0, 0.75, 0.0}
\definecolor{arsenic}{RGB}{50,30,90}
\definecolor{ao(english)}{rgb}{0.0, 0.5, 0.0}

\usepackage{xparse}
\usepackage{cleveref}

\title{\textsc{Gauge-underdetermination and shades of locality in the Aharonov-Bohm effect}\footnote{Preprint---accepted for \textit{Foundations of Physics} (Springer). Please cite published version, if available.}}

\author{Ruward A. Mulder\footnote{Freudenthal Institute, History and Philosophy of Science Department, Utrecht University, Princetonplein 5, 3584CC Utrecht, The Netherlands. \Letter  \, RuwardArthur@gmail.com; ram202@cam.ac.uk.}}
\date{}
\begin{document}
\renewcommand{\thefootnote}{\arabic{footnote}}
\maketitle

\begin{abstract}
\noindent I address the view that the classical electromagnetic potentials are shown by the Aharonov-Bohm effect to be physically real (which I dub: `the potentials view’). I give a historico-philosophical presentation of this view and assess its prospects, more precisely than has so far been done in the literature. Taking the potential as physically real runs \textit{prima facie} into `gauge-underdetermination': different gauge choices represent different physical states of affairs and hence different theories. This fact is usually not acknowledged in the literature (or in classrooms), neither by proponents nor by opponents of the potentials view. I then illustrate this theme by what I take to be the basic insight of the AB effect for the potentials view, namely that the gauge equivalence class that directly corresponds to the electric and magnetic fields (which I call the Wide Equivalence Class) is too wide, i.e., the Narrow Equivalence Class encodes additional physical degrees of freedom: these only play a distinct role in a multiply-connected space. There is a trade-off between explanatory power and gauge symmetries. On the one hand, this narrower equivalence class gives a local explanation of the AB effect in the sense that the phase is incrementally picked up along the path of the electron. On the other hand, locality is not satisfied in the sense of signal locality, viz. the finite speed of propagation exhibited by electric and magnetic fields. It is therefore intellectually mandatory to seek  \textit{desiderata} that will distinguish even within these narrower equivalence classes, i.e. will prefer some elements of such an equivalence class over others. I consider various formulations of locality, such as Bell locality, local interaction Hamiltonians, and signal locality. I show that Bell locality can only be evaluated if one fixes the gauge freedom completely. Yet, an explanation in terms of signal locality can be accommodated by the Lorenz gauge: the potentials propagate in waves at finite speed. I therefore suggest the Lorenz gauge potentials theory---an even narrower gauge equivalence relation---as the ontology of electrodynamics.
\end{abstract}

\clearpage


{\footnotesize{\hypersetup{linkcolor=black}\vspace{-4cm} \textsf{\tableofcontents}}}

 \section{Preamble: What there \textit{is} in the lecture halls}
 \noindent How do our theories of physics relate to what actually exists in the world? In the everyday life of physicists, both in the lab of the experimenter and on the paper of the theorist, such a question is rarely helpful for directly solving specific problems. Nevertheless, questions like `what \textit{is} the quantum state exactly?', `is there \textit{really} energy in the world?' or `does light \textit{actually} consist of photons or of electromagnetic waves?' are commonplace in the minds of students. Often, such questions are evaded, as if asking for things that are not obscure by accident and into which one should not delve too deeply. Unmistakeably, part of this reluctance about dealing with such questions lies in the difficulty of conceptual reasoning in mathematical theories that are designed to solve practical problems.   

In lectures on electrodynamics during undergraduate years in physics, the introduction of the electromagnetic potentials as follows is commonplace:
\begin{quote}
The fact that Maxwell's equations show that the electric and magnetic fields can propagate on their own and carry energy over long distances, shows that they are physically real. Let us now introduce the electromagnetic potentials through their relation with these fields, $\textbf{E} = - \nabla \phi -  \partial \textbf{A} / \partial t$ and 
$\textbf{B}=\nabla \times \textbf{A} $. These are introduced for mathematical convenience, mind you, and have more degrees of freedom than the actual physical degrees of freedom. 
\end{quote}
This appears unproblematic: the `actual physical' degrees of freedom are encoded in the $\textbf{E}$ and $\textbf{B}$ fields and the role of the `extra' structure can be understood as convenient for the practical purposes of the calculations that await the student. After all, the task of finding the value of the potential between two conducting plates that is high enough to ionize the hydrogen atom might not leave much time to question the physical nature of that voltage. 

But what is to stop an eccentric from adhering to a `potentials view', where $\phi$ and $\textbf{A}$ correspond to real things in the world? Empirical data does not rule out such a move, since the potential theory is empirically equivalent (at least in the classical domain). Of course, the term `Ockham's razor' (which is often bandied about) applies in the sense that we presumably should not add extra structure when it is \textit{unnecessary}. Yet opinions over what is necessary diverge. Particularly for these two electromagnetic theories---apparently making the same predictions---additional criteria come into play. 
 
Near the end of the semester,  the professor admits that his initial claim, that the potentials merely serve an auxiliary role, has a caveat:
\begin{quote}
Remember the claim that $\phi$ and $\textbf{A}$ are merely convenient constructs that only help us solve problems? It turns out that in quantum theory this is no longer really true, because of the Aharonov-Bohm effect. I will not go into it now, but you will see that gauge degrees of freedom have a direct effect on quantum observables.
\end{quote}
Although it makes for an exciting cliff-hanger, the student is often left hanging there. When the AB effect is discussed in the subsequent advanced quantum mechanics courses, the reasons for taking $\phi$ and $\textbf{A}$ to be physically real, and the nature of gauge degrees of freedom, are often omitted. Indeed, although some sense of locality is the sole reason for taking $\phi$ and $\textbf{A}$ as real, the word `locality' is rarely mentioned.

\section{Outline of the problem and the argument}
The Aharonov-Bohm effect shows that there are situations in which the phase of the wavefunction of a charged quantum probe is influenced at locations where the magnetic (and electric) field vanishes, whereas the vector potential $\textbf{A}$ does not. Thus, in brief, one must conclude that either the magnetic field $\textbf{B}$ acts  in a non-local way  or the $\textbf{A}$-field plays a physical role.\footnote{This dichotomy of course assumes that there is no third candidate that provides a mechanism to explain the effect but that is hidden somewhere in the formalism and therefore overlooked; cf. footnote \ref{VaidmanACR}.} But if the former: in what way? And if the latter: in what way? The Aharonov-Bohm effect (AB effect) thereby highlights the difficulty of understanding theories that admit local gauge transformations; it puts a question mark on our understanding of what gauge transformations are in the first place.

This paper is a historico-philosophical evaluation of the `potentials view' as opposed to the `fields view'. On the potentials view---adhered to by most physicists---the potentials, rather than the $\textbf{E}$ and $\textbf{B}$ fields, are taken as the physical players of the theory, so that the potentials are not just auxiliary mathematical conveniences.

In experimental physics, aspects of the AB effect have flourished in recent years (for example the occurrence of a magnetic edge in graphene rings \cite{Dauber}), but progress has been slower in finding an \textit{explanation} of the AB effect, as pointed out by Batelaan and Tonomura in 2009 \cite[p.38]{Batelaan} ``the investigation and exploitation of the AB effect remain far from finished." Tran recently challenged the idea that the formulation of Maxwell's equations is settled science, even in the classical case \cite{Tran}.  Although alternative ways of deriving the effect in didactic ways are promoted, for example by focusing on the de Broglie wavelength \cite{Kasunic}, the conceptual consequences of the potentials view are rarely taught. Berry adopts the potentials view by claiming that the ``gauge-invariant part of the vector potential was promoted to a real physical field, not just a convenient device for summarizing certain information about the electric and magnetic  fields" \cite{Berry}.

 To make things more precise, let me define the two views that involve an interpretation of the potentials.
\begin{quote}
\underline{Potentials view}: the scalar potential field $\phi(\textbf{x})$ and vector potential field $\textbf{A}(\textbf{x})$ are physical (that is, more real than mathematical fictions), over and above the degrees of freedom that give rise to the electric field $\textbf{E}(\textbf{x})$ and magnetic field $\textbf{B}(\textbf{x})$. A short-hand notation for this interpretation will be ($\phi,\textbf{A}$)-theory.
\end{quote}
This is in contrast with the rival view:
\begin{quote} 
 \underline{Fields view}: only the degrees of freedom that directly give rise to the electric field $\textbf{E}(\textbf{x})$ and magnetic field $\textbf{B}(\textbf{x})$ are  physical and any additional (mathematical) degrees of freedom of the potentials are pure mathematical fictions.  A short-hand notation for this interpretation will be ($\textbf{E},\textbf{B}$)-theory.
 \end{quote}
The fields view is the traditional interpretation, but it is logically quite strong: its defenders  share the belief that the electric and magnetic fields are sufficient for explaining all electrodynamic phenomena. On the contrary, within the potentials view there is ample leeway. The reason is that one has to choose where to draw the line between physical and fictional degrees of freedom, resulting in many different positions---none of which are uncontroversial\footnote{Here I give a brief review of prominent interpretations of electrodynamics in light of the AB effect. The first is the holonomies view, advocated by Healey, in which the loop integrals of the potentials are promoted to physical ontology \cite{HealeyBook}. Unlike the potentials, these loops are gauge-invariant quantities; unlike the fields, they are non-locally possessed properties which nonetheless act locally---and at least they are defined in the region where the electrons move \cite{Myrvold}. Another view is taken by Mattingly, who argues that gauge fields do not commit us to any novel ontology and that the effect should be understood in terms of the sum of 4-current fields of single charged particles \cite{Mattingly1}. Here, `distributivity' fails in the sense that the electron is sensitive to the component fields, even while the net field vanishes. The 4-current field would carry information but no energy-momentum \cite{Mattingly2}. Boyer argues that the AB effect is explained by the back-reaction of the \textbf{E} and \textbf{B} fields of the charged particles themselves, which interact with the solenoid \cite{Boyer2}. Vaidman argues that the potentials merely \textit{seem} real due to the stringent canonical formulation of quantum dynamics (the Schr\"odinger equation \textit{necessarily} deals with a potential instead of fields or forces: cf. Eq. \eqref{Hamil}). Therefore, Vaidman calls upon the community to seek a reformulation of quantum mechanics in terms of the electric and magnetic fields and without potentials \cite{Vaidman2012,Vaidman2015}.  Aharonov himself, together with Cohen and  Rohrlich replied to Vaidman that the effect may be due to a local gauge potential or due to non-local gauge-invariant fields \cite{ACR2015,ACR2016}. Clarity on these issues is especially important in the light of the calculations performed by Pearle and Rizzi, who have worked out Vaidman's idea of including the solenoid in the AB experiment into a fully quantum-mechanical description \cite{Rizzi}. This `Vaidman-ACR debate', as we might dub it, is a contemporary example of a dispute in which extra-empirical values play a crucial role in physical practice. I intend to evaluate this particular debate in future work.\label{VaidmanACR}}---but which share commitment to the physicality of all or some of the degrees of freedom which were considered unphysical degrees of freedom by the traditional fields view. 

In what follows, I will initially be concerned with the naive version of the potentials view which says `the potentials are real' without specifying what we mean by that. This will involve some fuzziness about both locality and reality criteria; but the initial focus on the naive view helps to tease out the philosophical commitments that play a role in these controversies. Also, this view appears to me to be the most straightforward way to give a local explanation of the AB effect. In the second half of the paper, the focus will shift to the more precise language of equivalence classes in order to sharpen intuitions about locality and ontology.

I will approach the debate over the ($\phi$,\textbf{A})-theory or (\textbf{E},\textbf{B})-theory from the viewpoint of underdetermination, which sheds light on the role of locality and reality in theory choice. I explore the Aharonov-Bohm effect and the potentials formalism in sections~\ref{secAB}-\ref{fieldspotentials} in such a way that all steps in the calculations are present. Then, I emphasize that if one regards the potentials as completely real, as many do, there must also be a preferred gauge. If one does not  choose a preferred gauge, it is impossible to avoid what I call `gauge-underdetermination' (section \ref{GU}). Thus, one  needs either to appeal to a criterion for the preferred gauge or to weaken the statement that the potentials are real. This is often brushed over or misunderstood in the literature about gauge theories, as illustrated by excerpts from influential authors (section \ref{absence}), sometimes leading to inconsistencies in the treatment of the AB effect. 

\begin{figure}
\centering
\includegraphics[width=1.1\textwidth]{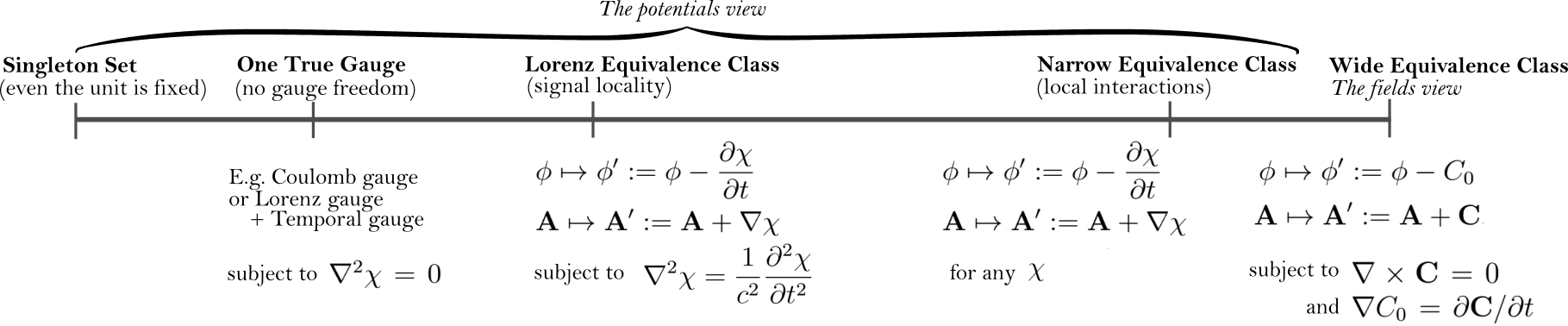}
\caption{\footnotesize{A spectrum of equivalence relations on the potentials and corresponding philosophical positions with wider relations towards the right. On the far right, one has the fields view for which only  $\textbf{E}$ and $\textbf{B}$ are real. Everything to the left of this posits more physical degrees of freedom and is hence within the potentials view. The Narrow Equivalence Class, which is the one most often used in contemporary physics, encodes fewer gauge degrees of freedom---it coincides with the fields view in simply-connected spacetimes. One can narrow this down further by additional constraints on the local gauge function $\chi$, such as my suggested Lorenz gauge. Fixing \textit{all} degrees of freedom, for example in  Coulomb gauge, would lead to Maudlin's `One True Gauge' potentials theory. For completeness: such a position would still involve a residual but trivial equivalence class, reflecting possible choices of units. The reader is invited to find additional positions.}} \label{fig:spectrum}
\end{figure}

 To find criteria that soften gauge-underdetermination, one must recognize that  \textit{cutting down on gauge-underdetermination and narrowing one's gauge equivalence class are two sides of the same coin}. To do this, it is natural (but not necessary) to remain close to the original motivation: locality. In section \ref{Locality}, I survey several precise formulations of a locality condition, such as Bell locality, local interaction Hamiltonians, separability and signal locality. In section \ref{Classes}, I review the reasoning that led Aharonov and Bohm to infer the reality of the potentials. I discuss how this can be read as narrowing down the `old' Wide Equivalence Class of admitted gauge symmetries ($\textbf{A}\mapsto\textbf{A}':=\textbf{A}+\textbf{C}$ such that $\nabla \times \textbf{C}=0$) to the Narrow Equivalence Class ($\textbf{A}\mapsto\textbf{A}':=\textbf{A}+\nabla \chi$), so as to admit some physical degrees of freedom of the potential in addition to those that give rise to $\textbf{E}$ and $\textbf{B}$. The (modern) Narrow Equivalence Class is local in a sense, namely that the electron picks up a phase incrementally along its path. The kind of locality I then suggest we should seek is signal locality,  similar to the locality exhibited by $\textbf{E}$ and $\textbf{B}$, since they propagate with a finite velocity. With that knowledge, I suggest that one can go a step further and extend this narrowing of the gauge equivalence class: I argue in section \ref{Calc} that the `Lorenz Equivalence Class' satisfies that \textit{desideratum}.

The overall argument of the paper thus consists of there being several options on a spectrum of gauge equivalence classes, as shown in Figure \ref{fig:spectrum}---according to one's commitments to locality and reality, one can  explore additional positions to those shown.

  \section{Aharonov-Bohm: non-locality and gauge potentials} \label{secAB}
Due to their prediction of the effect that came to carry their names,  David Bohm and his PhD student Yakir Aharonov \cite{AB} argued that a ``further interpretation of the potentials is needed in quantum mechanics." Their solution, which was adopted by many physicists, was to promote the potentials from mathematical fiction to something physical.\footnote{This effect had been semi-classically calculated by Ehrenberg and Siday in 1949 \cite{ES}, and might therefore be called the Aharonov-Bohm-Ehrenberg-Siday effect. Aharonov and Bohm clearly stressed the metaphysical importance of this result, which seems to be the main reason why the phenomenon carries their name.}

\begin{figure}
\centering
\includegraphics[width=0.65\textwidth]{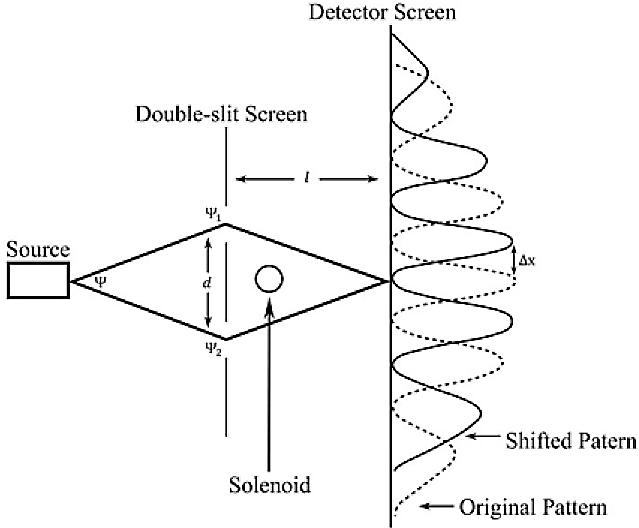}
\caption{\footnotesize{Experimental set-up to show the AB interference effect with time-independent vector potential as a result of switching on the current through the solenoid. The usual two-slit interference pattern is translated upward by an amount $\Delta x = - q\lambda_{B}l\Phi/2 \pi \hbar d $. Figure from \cite[p.7]{Ardourel}.}} \label{fig:1}
\end{figure}

As illustrated in Fig.~\ref{fig:1}, Aharonov and Bohm envisage the following experimental set-up, which originated as a thought experiment but was shortly afterwards experimentally realized by Robert Chambers (who is mentioned in their original paper). A coherent electron beam is split and directed around a solenoid and brought together again in a region where interference can be detected at a screen. The solenoid can be imagined as an infinite tightly-wound coil (so that the current is strictly circular, without a perpendicular component), which in turn ensures, by the usual magnetostatic symmetry arguments, that the magnetic field is \textit{confined} to inside the solenoid (in the direction perpendicular to the plane of the figure). The solenoid itself is then shielded, so that the wavefunction of the electrons is excluded from the region occupied by the solenoid. 

In a canonical formulation, external magnetic fields are coupled to the electron via the electromagnetic potential. The Hamiltonian  for an electron of mass $m$, momentum $\textbf{p}$  and charge $q$ in an external magnetic field is
\begin{equation} \label{Hamil}
H=\frac{1}{2m}\brack{\textbf{p}-q\textbf{A}}^2,
\end{equation}
where $\textbf{A}$ is stationary \cite{Peshkin}. If the electron wavefunction in the absence of the magnetic field is given by $\psi_0(\textbf{x})$, the presence of the magnetic field forces us to add a phase to the electron wavefunction---the Dirac phase factor \cite{Dirac}---calculated by solving the time-independent Schr\"odinger equation,
\begin{equation}\label{electron}
\psi(\textbf{x})=\psi_0(\textbf{x})\exp\brack{\frac{i q}{\hbar}  \int \textbf{A} \cdot d\textbf{x}},
\end{equation}
where the phase factor is given by the line integral of the vector potential over some path in the  region where the electron is allowed to move (that is, it can take any path outside the solenoid).\footnote{In the four-vector formalism one can accommodate the time-dependent version of the AB effect with the Dirac phase factor $ \brack{q / \hbar} \oint_{\gamma} A_{\mu}  dx^{\mu} = \brack{q / \hbar} \oint_{\gamma} \blockbrack{\textbf{A} \cdot d\textbf{x} - \phi dt} $ over some closed path $\gamma$ through spacetime. I will steer clear of using covariant notation to \textit{(i)} keep the discussion accessible to the undergraduate level, which it conceptually surely is, and \textit{(ii)} to prevent the activation of already-possessed knowledge of the experienced gauge-theorist, particularly the tendency to readily interpret only gauge-invariant quantities as physical without reconsidering which symmetry it is based on.}

Decomposing the wavefunction of the electron into a superposition of two parts, $\psi(\textbf{x})=\psi_1(\textbf{x})+\psi_2(\textbf{x})$, one describing the electron following a path $1$ in the clockwise direction around the solenoid and the other the counter-clockwise path $2$, the phase difference between these paths in the interference region is
\begin{equation}
\frac{\Delta S}{\hbar} = - \frac{q}{\hbar} \brack{ \int_{\text{path 1}} \textbf{A} \cdot d\textbf{x} - \int_{\text{path 2}} \textbf{A} \cdot d\textbf{x} } = -  \frac{q}{\hbar} \oint_C \textbf{A} \cdot d\textbf{x},
\end{equation}
which is a line integral over the closed path $C$ defined by traversing path $1$ and then, in reverse, path $2$. The two phases picked up along the paths do not cancel, but add to give the loop integral because of the opposite orientations of the integrals of the two paths.  

But what fixes the path of integration so that we can actually evaluate the integral? After all, a quantum particle can be regarded as taking  all possible paths. This is rather subtle because the value of $\textbf{A}$ is subject to a gauge condition. The approach is as follows. For some closed loop $\gamma$, using the \textit{familiar but tailor-made} gauge freedom $\textbf{A} \mapsto \textbf{A}' = \textbf{A} + \nabla \chi$, we find the following equality
\begin{equation}\label{vanish}
 \oint_\gamma \textbf{A}' \cdot d\textbf{x}=\oint_\gamma \brack{\textbf{A}+\nabla \chi} \cdot d\textbf{x}=\oint_\gamma \textbf{A} \cdot d\textbf{x},
\end{equation}
where the second equality holds due to the identical vanishing of a gradient around a closed path. Eq. \eqref{vanish} shows that the loop integral  $\oint_\gamma\brack{\textbf{A}'-\textbf{A}}\cdot d\textbf{x}=0$ vanishes.\footnote{To visualize this better, we can, following Binney and Skinner \cite[Sec. 3.3.3]{BinneySkinner}, consider the form of the vector potential for an infinitely thin solenoid (at the origin and in the $\hat{z}$-direction) in cylindrical polar coordinates, so that $\textbf{A}= \frac{\Phi}{2\pi r^2} \brack{-y,x,0}$ for $r=\sqrt{x^2+y^2}$ and $\Phi$ the magnetic flux (cf. Eq. \eqref{flux}). Now choose the gauge such that this vector field vanishes, namely   $\nabla \chi = \frac{\Phi}{2\pi r^2} \brack{y,-x,0}$, which is achieved by choosing $\chi=-\frac{\Phi}{2\pi}\theta$ for the polar angle $\theta=\arctan\brack{y/x}$.\label{Binney}}  In other words, the line integral over $\int_{\textbf{x}_1}^{\textbf{x}_2} \brack{ \textbf{A}'-\textbf{A}} \cdot d\textbf{x}$ depends only on the initial position $\textbf{x}_1$ and final position $\textbf{x}_2$, not on the path taken. The reason why I call this gauge freedom tailor-made is that it  is specifically designed to make this integral vanish, making it narrower than a more general gauge symmetry  $\textbf{A} \mapsto \textbf{A}' = \textbf{A} + \textbf{C}$ with $\nabla \times \textbf{C}=0$, as I will discuss in section \ref{GU}.

What this calculation implies is a shift in the interference pattern depending on the magnetic vector potential $\textbf{A}$. When the distance $l$ between the screen and the slits is much larger than the size of the screen in the interference region (given by the position variable $x$), the usual two-slit interference pattern, $\Delta S_{\text{two  slit}}/\hbar=2\pi x d / \lambda_{B}l$ (where $\lambda_{B}=h/|\textbf{p}|$ is the de Broglie wavelength of the electron and $d$ is the separation between the slits),  shifts upward (or downward, depending on the direction of the current) by an amount\footnote{The early experiments by Chambers and by Moellenstedt and Bayh did not involve slits. Also, Timothy Boyer has kindly pointed out to me that Figure \ref{fig:1} and my phrasing incorrectly suggests that the whole interference pattern is displaced sideways undisturbed, whereas in reality the double-slit interference pattern changes, not the single-slit envelope \cite{Boyer}: an inaccuracy or lack of nuance that he traces back to the \textit{Feynman Lectures}. Here, it is sufficient to understand the problem through Figure \ref{fig:1}.} 
\begin{equation} \label{shift}
 \Delta x = \frac{ \lambda_{B}l}{2 \pi d} \frac{\Delta S}{\hbar} = -\frac{ \lambda_{B}l}{2 \pi d} \frac{q}{\hbar} \oint_C \textbf{A} \cdot d\textbf{x}= -\frac{ \lambda_{B}l}{2 \pi d} \frac{q}{\hbar} \Phi.
\end{equation}
 To identify $\Phi$ as the magnetic flux through the region that is enclosed by the loop $C$, one uses Stokes' theorem to rewrite this phase shift in terms of the magnetic field $\textbf{B}$ through surface area $S$ with the loop $C$ as its boundary,
\begin{equation}\label{flux}
 \oint_C \textbf{A} \cdot d\textbf{x} = \int \brack{ \nabla \times \textbf{A} } \cdot d\textbf{S} = \int  \textbf{B} \cdot d\textbf{S} := \Phi .
\end{equation}
That we can write the loop integral in terms of the magnetic field is not a remarkable result by itself, since it is, after all, expected that magnetic fields influence electrons' behaviour. The remarkable fact is that \textit{the magnetic field is confined to the region inside the solenoid and is zero outside}. The electrons experience a shift even though they never move through a region of non-zero magnetic field: but the shift still depends on the current.

Aharonov and Bohm argue that because ``in a field-free multiply-connected region of space, the physical properties of the system still depend on the potentials," we should  promote the potentials from fiction to being physical:
\begin{quote}
[t]he Lorentz force $\blockbrack{e\textbf{E}+\frac{e}{c}\textbf{v} \times \textbf{B}}$ does not appear anywhere in the fundamental theory, but appears only as an approximation holding in the classical limit. It would therefore seem natural at this point to propose that, in quantum mechanics, the fundamental physical entities are the potentials, while the fields are derived from them by differentiations \cite[p.490]{AB}.
\end{quote} 
The important point is that Aharonov and Bohm appeal to some kind of explanation of the AB phase \textit{using the tools at hand} in the mathematical formalism of the theory. Even if one is in principle willing to allow a non-local influence of the B-field on the electron,\footnote{Indeed, David Bohm himself would not be so bothered by non-local explanations in physics.} the structure of that B-field says nothing about how this non-local influence would work. After all, the B-field satisfies the wave equation so that it propagates at a finite speed $c$, so that it surely cannot be invoked  as acting non-locally.\footnote{In the jargon I introduce in section \ref{Locality}, $\textbf{B}$ satisfies \textit{signal locality}, which is here fulfilled by satisfying the wave equation (given that $\textbf{B}$ is physical).} Hence, the (\textbf{E},\textbf{B})-theory puts no explanation of this phenomenon on the table.

Hence, the AB effect directly challenges the fields view. The predictions of the combination of classical (\textbf{E},\textbf{B})-theory and quantum-mechanical test particles fail, forcing a reconsideration in either one of these theories. One option would be to search for a non-local law in quantum mechanics, which at first sight sounds reasonable in view of the violation of the Bell inequalities, as will be discussed in section \ref{Locality}. The other option, which is widely adopted and will now be explored in more detail, is to follow Aharonov and Bohm's advice in promoting the potentials to `fundamental ontology'.

\section{Electrodynamics in terms of potentials} \label{fieldspotentials}
It makes a difference which parts of the  mathematical formalism of a theory are seen as physical. Two theories that share the same formalism can differ in their assessment of the physicality of part of that formalism. Those parts that are supposed to correspond (in some admittedly philosophically controversial way\footnote{According to Quine, the ontology is given by the domain you quantify over \cite{Quine}; closest to that in spirit is a primitive ontologist who postulates an explicit fundamental ontology together with axioms \cite{Allori}; structural realists say that the ontology is provided by the mathematical structures used in the axioms \cite{Ladyman}. The language in this paper is naturalistically inclined: I will follow most physicists in labelling those objects that are supposed to be in the world as `physical'.\label{link}}) to entities in the actual world, I will call `fundamental ontology': the physical building blocks according to the theory (I say `according to the theory' to prevent the connotation of `ultimate building blocks', i.e., I take fundamentality as a relation between  theories). Yet, regardless of the way the relation between theory and the world is fleshed out, it is clear that it is entirely possible---even the historical norm---that what parts of the formalism are considered to be physical becomes contested in times when a new theory is needed. 

In electrostatics, the electric field can be interpreted as merely encoding the  propensity of a test charge to accelerate under the (non-local) Coulomb force. In electrodynamics, Maxwell's equations  do not deal directly in terms of forces, but with the fields and the source charges. Electrodynamical phenomena can be described by Maxwell's equations, which, in vacuum, read: Gauss' law, $\nabla \cdot \textbf{E}=0$; Faraday's law,
$\nabla \times \textbf{E} = -  \partial \textbf{B}/ \partial t$;
Gauss' law for magnetism,
$\nabla \cdot \textbf{B}=0$;
and the Maxwell-Amp\`ere law,
$\nabla \times \textbf{B} = (1/c^2) \partial \textbf{E} / \partial t$.  These equations lead\footnote{Taking the curl of Faraday's law, using the vector identity $\nabla \times (\nabla \times \textbf{a})=\nabla (\nabla \cdot \textbf{a}) - \nabla^2 \textbf{a}$, invoking the constraint that is Gauss' law and using the Maxwell-Amp\`ere law.\label{VI}} to electromagnetic waves with speed $c$, which can travel for millions of years in the absence of any nearby charges. These fields carry energy and momentum, as expressed by the Poynting vector $\textbf{S} \propto \textbf{E} \times \textbf{B}$, which gives the energy flux of an electromagnetic wave.  This leads one to interpret the electric and magnetic fields as fundamental ontology, as opposed to a mere propensity of a test charge to accelerate when subject to the Coulomb force. This interpretive move is made in every course on electrodynamics. Besides, here I should mention the unification of optics and electromagnetism---and it is this explanation that convinces the student that there is something more to the vector fields $\textbf{E}$ and $\textbf{B}$ than meets the eye: light.

We can attempt a similar step to favour the ($\phi$, \textbf{A})-theory over the   (\textbf{E},\textbf{B})-theory. An alternative formulation in terms of the scalar potential $\phi$ and vector potential $\textbf{A}$ adds some extra structure to the fields, from which the field can be obtained via the equations 
\begin{equation} \label{potential1}
\textbf{E} = - \nabla \phi - \frac{\partial \textbf{A}}{\partial t}, 
\end{equation}
and
\begin{equation} \label{potential2}
\textbf{B}=\nabla \times \textbf{A}.
\end{equation}
These definitions are chosen such that Gauss' law of magnetism and Faraday's law are trivially satisfied. With these definitions, the two dynamical equations (the time-derivatives in Faraday's law and the Maxwell-Amp\`ere's law encode the evolution of the fields) become
\begin{equation}
\nabla^2 \phi  =- \frac{\partial}{\partial t} \brack{\nabla \cdot \textbf{A} }, \label{GaussPot} 
\end{equation}
and (using the same vector identity as in footnote \ref{VI})
\begin{equation}
\nabla ^2 \textbf{A} - \frac{1}{c^2} \frac{\partial^2 \textbf{A}}{\partial t^2} = \nabla\brack{\nabla \cdot \textbf{A} + \frac{1}{c^2} \frac{\partial \phi }{\partial t}}  \label{MaxAmpPot}.  
\end{equation} 
\noindent The form is less neat than in the (\textbf{E},\textbf{B})-theory, but this shows that a formulation in terms of potentials is possible without making reference to the electric and magnetic fields at all.  

Both formulations are empirically equivalent since their predictions---at least in the known domains that have been  probed---will not depend on any of the `extra' structure introduced by the definitions Eqs. \eqref{potential1}-\eqref{potential2}. The reason is that the values of the electric and magnetic fields we can measure, and the trajectories of charged particles in these fields as determined by the Lorentz force law, remain the same regardless of the choice between the (\textbf{E},\textbf{B})- or ($\phi$,\textbf{A})-theories. 

Here is a clear case of what philosophers call `underdetermination of theory by data'. So the following position can be maintained: an eccentric can claim that we should postulate the potentials $\phi$ and $\textbf{A}$ as directly related to the `real' things in the world, instead of the electric and magnetic fields. That position is the most straightforward version of the potentials view. Note that the (\textbf{E},\textbf{B})-theory and the ($\phi$,\textbf{A})-theory are regarded as two \textit{different} \textit{theories} because they have different fundamental ontologies. In the real world, including special circumstances in the lab, we see the potentials $\phi$ and $\textbf{A}$  just as indirectly as the fields $\textbf{E}$ and $\textbf{B}$, although we might intuitively feel more comfortable with the latter. 

 On merely empirical grounds the case must remain undecided as long as one cannot give an independent reason to take the potentials seriously as real things. But the Aharonov-Bohm effect provides precisely such an independent reason, since the ($\phi$,\textbf{A})-theory provides \textit{local} explanations of the AB phase shift in Eq.~\eqref{shift}, whereas the (\textbf{E},\textbf{B}) theory cannot give such an explanation. This is why many endorse the potentials view. In the next section I will argue that, if one is to take the potentials view, then further steps have to be considered in order to have a well-defined theory---a fact that often goes unmentioned.

 \section{`Gauge-underdetermination' and equivalence classes}\label{GU}
 Gauge symmetry is usually formalized by the transformations 
\begin{equation}\label{gauges}
\phi \mapsto \phi' := \phi -\frac{\partial \chi}{\partial t},   
\end{equation}
and
\begin{equation}\label{gauges2}
\textbf{A} \mapsto \textbf{A}' := \textbf{A} + \nabla \chi,  
\end{equation}
where $\chi(\textbf{x},t)$ is an arbitrary scalar function, depending on both space and time coordinates. The fields $\textbf{E}$ and $\textbf{B}$ are left unchanged regardless of the function $\chi$, thanks to the definitions \eqref{potential1} and \eqref{potential2}. Hence, if one is only interested in $\textbf{E}$ and $\textbf{B}$, one can choose a $\chi$ that simplifies the derivation of a solution to a given physical problem. In principle, an infinity of gauges is possible, as the scalar function $\chi$ is arbitrary.

For example, Maxwell's equations in the potentials formulation can be given a simpler form by choosing the Coulomb gauge (often used in magnetostatics),
 \begin{equation} \label{CoulombGauge}
\nabla \cdot \textbf{A} = 0.
\end{equation} 
In other words, the gauge parameter $\chi$ is constrained by $\nabla^2 \chi = 0$, which has a unique solution and is therefore a complete gauge-fix.  Gauss' law \eqref{GaussPot} and the Maxwell-Amp\`ere law \eqref{MaxAmpPot} then become
\begin{equation}
\nabla^2 \phi  = 0, \label{GaussPotCoulomb}
\end{equation} 
\begin{equation}
\nabla ^2 \textbf{A} - \frac{1}{c^2} \frac{\partial^2 \textbf{A}}{\partial t^2} = \frac{1}{c^2} \frac{\partial }{\partial t}\nabla\phi.   \label{MaxAmpPotCoulomb}
\end{equation}
Eq.~\eqref{GaussPotCoulomb} is simply the Laplace equation, but the Maxwell-Amp\`ere law has mixed space- and time-derivatives. 

Alternatively, if one were to choose the Lorenz gauge,\footnote{Not Loren\textit{t}z, although often  this choice of gauge is connected to Hendrik Antoon Lorentz instead of Ludvig Lorenz: a confusion well documented by van Bladel \cite{Bladel}.} it precisely cancels that mixed term, by requiring
\begin{equation}\label{LorenzGauge}
 \nabla\cdot \textbf{A} + \frac{1}{c^2}\frac{\partial \phi}{\partial t} =  0.
\end{equation}
In other words, the gauge parameter $\chi$ is constrained by $\nabla^2 \chi = 1/(c^2)\partial^2/dt^2$, which is an incomplete gauge condition, as there is residual freedom in the form of a scalar wave (one can choose a further fix that leaves no residual freedom, for example by setting $\phi=0$, which is the so-called  temporal gauge).  In the Lorenz gauge, Eqs.~\eqref{GaussPot}~and~\eqref{MaxAmpPot}  reduce to
\begin{equation} \label{GaussPotLorenz}
\nabla^2 \phi = \frac{1}{c^2} \frac{\partial^2 \phi}{\partial t^2}
\end{equation} 
and
\begin{equation} \label{MaxAmpPotLorenz}
\nabla^2 \textbf{A} = \frac{1}{c^2} \frac{\partial^2 \textbf{A}}{\partial t^2},
\end{equation}  
\noindent resulting in the wave equation for both the vector and scalar potentials. There is also a residual gauge degree of freedom, which can be fixed by an initial condition that specifies $\phi$ at spatial infinity---which I discuss further in section \ref{Calc}.

 Equations \eqref{gauges}-\eqref{gauges2} define an equivalence class. Thus it is said that any choice of $\chi$ will lead to the same physics. But one has to first agree on what the physics is. In this case, the usual argument is that this equivalence class leaves the electric and magnetic fields invariant. This is of course true. Yet, anticipating section \ref{Classes} and turning to what I called (just below Eq. \eqref{vanish}) the tailor-made gauge symmetry, there is, in fact, a wider equivalence class one can consider which leaves the fields invariant. Namely, for any $\textbf{C}$ and $C_0$ such that $\nabla \times \textbf{C}=0$ and $\nabla C_0 = \partial \textbf{C}/ \partial t$, one transforms
\begin{equation} \label{gaugevector}
\phi \mapsto \phi':=\phi-C_0, 
\end{equation}
and
\begin{equation} \label{gaugevector2}
\textbf{A} \mapsto \textbf{A}':=\textbf{A}+\textbf{C}.
\end{equation}

 This defines an equivalence class of potentials, related to each other by the choice of gauge vector $\textbf{C}$ and gauge scalar $C_0$. This class is \textit{wider} than the class defined by Eqs. \eqref{gauges}-\eqref{gauges2}, where the gauge freedom is given only by the possible choices of the scalar $\chi$. There is more freedom in the gauge vector $\textbf{C}$ than in the gauge scalar function $\chi$, even though both classes give the \textit{same} electric and magnetic fields (for Eq. \eqref{potential2} due to the imposed requirement   that the rotation of $\textbf{C}$ vanishes and for Eq. \eqref{potential1} because the rotation of a gradient vanishes identically).
 
  Therefore, I will speak of the \textit{Wide Equivalence Class} defined by \eqref{gaugevector}-\eqref{gaugevector2} and the \textit{Narrow Equivalence Class} defined by  \eqref{gauges}-\eqref{gauges2}.  Agreed: there is no difference between the two classes in a simply-connected space. Yet there are no local facts that allow  us to figure out how our space is connected. It is precisely the legacy of Aharonov and Bohm, whose thought experiment introduces a multiply-connected space (the solenoid `punctures' the space, as it were, introducing a conical singularity), that one now works with the Narrow Equivalence Class, which provides  additional degrees of freedom that can \textit{in some sense} account for the AB phase shift. Section \ref{Classes} is devoted to what sense this is. This relates to the goal of this paper: to shed light on the path of narrowing this class down even further to obtain a clear local ontology. 

Back to the naive potentials view: the ``eccentric" who wants to hold on to regarding $\phi$ and $\textbf{A}$ as completely real cannot claim that a choice of gauge `does not change the physics'. For her, the potentials are the physics and, hence, every different $\chi$ corresponds to a different theory.   Each choice of $\chi$ leaves the fields $\textbf{E}$ and $\textbf{B}$ unchanged; and hence these fields \textit{underdetermine} the theory, on the potentials view. However, down this path one quickly encounters a straightforward problem. There is no independent way to determine the value of $\chi$, since it can no longer be seen as simply a pragmatic tool for the purposes of calculation. In other words, the predictions are ambiguous; the equations of motion will have non-unique solutions and one is left with an indeterministic theory.

On the potentials view, determinism manifestly fails as long as $\chi$ is not fixed. For one can make an argument analogous to the `hole argument' that exploits diffeomorphisms in general relativity \cite{Norton}. Suppose the fields view and the potentials view are related by a time-dependent gauge transformation that is the identity up to a time $t$. Hence, they agree up to $t$, but diverge afterwards. If different values of $\chi$   lead to different facts about the real world---which the fields view does not claim but the potentials view does---this amounts to indeterminism. For there are multiple ways that $\phi$ and $\textbf{A}$ can evolve in time.  

The ($\phi$,\textbf{A})-theorist should not want to endorse such  indeterminism---certainly not in response to an eminently predictable phenomenon such as the AB effect---since the theory would be unable to make definite predictions (although it could restrict the predictions to merely $\textbf{E}$ and $\textbf{B}$). Not even the option to assign probabilistic weights to the alternatives, that is used in textbook quantum mechanics, is available to her.  Therefore, the ($\phi$,\textbf{A})-theorist should embrace that Eqs.~\eqref{gauges}~and~\eqref{gauges2} represent a \textit{set} of theories, parametrized by $\chi$. Hence, the flipside of gauge-indeterminism is what I call `gauge-underdetermination'. But this underdetermination is only problematic if one leaves $\chi$ unfixed. 
    
 The conclusion then \textit{seems} unavoidable. If one claims that the potentials are completely physical, the potentials view leads to indeterminism: the exact same values of $\textbf{E}$ and $\textbf{B}$ can be predicted using different gauges which correspond to different futures of $\phi$ and $\textbf{A}$ given fixed initial conditions.  The way to solve this and recover the ability to make unique predictions  is to regard \textit{only one} of the gauge choices as genuinely physical. Maudlin\footnote{Tim Maudlin, incidentally, does not defend this position as his personal solution to the ontological problems posed by the AB effect. The position originated in a response to Healey's `mid-way' position about the reality of the potentials.} has called this the `One True Gauge' principle  \cite{Maudlin}.  Hence, it is not arbitrary to choose between the Lorenz, the Coulomb or some other gauge. Different physical facts corresponds to different choices of  gauge. Once a unique choice has been made and the theory accepted, the other gauge choices can be considered as mathematically convenient fictions, as tools for making calculations suited to different problems.    That is why the name `($\phi$,\textbf{A})-theory' is not sufficient and should really be seen as a collection of `($\phi$,\textbf{A})-theories', of which, for example, the Coulomb gauge potentials theory, ($\phi$,\textbf{A})$_{\text{CG}}$-theory, and the Lorenz gauge potentials theory, ($\phi$,\textbf{A})$_{\text{LG}}$-theory, are members.  
 
 We need not be so ambitious, however, as to consider \textit{all} degrees of freedom encoded in $\phi$ and $\textbf{A}$ physical. One can define gauge equivalence classes whose members all match each other on the degrees of freedom that one considers physical. The above `collection' of ($\phi$,\textbf{A})-theories thus corresponds to the Narrow Equivalence Class. Depending on one's criteria, one would narrow this equivalence class down further, which is equivalent to adding more constraints, i.e., shrinking the space of functions for $\chi$.  
 
 In section \ref{Locality}, I present candidates for trimming down gauge-underdetermination in the form of  Bell locality and signal locality. In section \ref{Classes}, I discuss Aharonov and Bohm's locality concept and modern views of the local explanation that can be given in terms of the equivalence class defined by Eqs. \eqref{gauges}-\eqref{gauges2}.  This class is suitable to provide an explanation in terms of local interactions, but is deemed insufficiently suited to provide a signal-local explanation of the AB effect in terms of a sharp ontology. Thus, I will formulate \textit{desiderata} that can lead to a further narrowing. In section \ref{Calc}, I will suggest that picking the Lorenz gauge as such a constraint will narrow the gauge equivalence class down in such a way that one has a signal-local explanation of the AB effect in terms of travelling $\phi$ and $\textbf{A}$ potential fields, just as one has travelling  $\textbf{E}$ and $\textbf{B}$ fields in the traditional Maxwell theory. But before that, I discuss in the next section how and why the considerations that go into choosing a gauge equivalence class, and in particular the related problem of gauge-underdetermination, are so often not engaged with in the scientific literature.

 \section{Gauge-underdetermination in the scientific literature} \label{absence}
 The fact that underdetermination automatically arises in the potentials view is  too often unacknowledged. In this section, I will criticise the presentations of several widely-read authors. In general, the issue is that even though these authors commit to the potentials being real, they are still considered as gauge fields where the gauge transformation remains as free a tool as in the fields view. This is having your cake and eating it too---using the potentials view for an explanation of the AB effect and the fields view to get out of gauge-underdetermination.  
 
 Norsen devotes a section of his recent (important) book to  questions about reality and locality similar to those in this paper. He considers the possibility of accepting the electrodynamic potentials as physically real \cite[p.21]{Norsen}. Then he alludes to Bell's argument (cf section \ref{Locality}) that the scalar potential is non-local in the Coulomb gauge and argues, like Bell, that this is not a problem because it is ``bound up in some way with human knowledge or conventions." 

But \textit{what} is bound up with human knowledge or convention? If it is the potentials, then we are simply rejecting the premise that the potentials are real. Hence it is more likely that Norsen means the gauge choice. But if one regards the potentials as physically real, the gauge choice can no longer be considered a convention, since every `choice' leads to a different description of the state of affairs in nature.

 Another instance is found in the introduction to gauge theories by Moriyasu, who states  that
\begin{quote}
[t]he Aharonov-Bohm effect clearly contradicted the accepted notion that only the electric and magnetic fields could produce observable effects. More important, it became evident that the potential had to be treated as a physical field that was also directly observable. The alternative would be to believe that the phase shift is produced by the magnetic field ``acting at a distance" in direct conflict with relativity \cite[p.21]{Moriyasu}.
 \end{quote}
But the precise way in which the magnetic field could act at a distance is not fleshed out. More importantly, there is no mention in the book about the consequential  underdetermination that arises from taking the potentials as the fundamental ontology; at the same time, the author speaks of making different gauge choices as if the fields view was adhered to. 
 
In his Lecture 9, Feynman defined locality as an intrinsic part of (real) fields: ``a field is `real' if it is what must be specified \textit{at the position} of the particle in order to get the motion" \cite{FeynmanLec}.  He further made the sociological observation that this conclusion had become consensus, since ``\textbf{E} and \textbf{B} are slowly disappearing from the modern expression of physical laws; they are being replaced by \textbf{A} and $\phi$." The latter was surely correct at the time and has become even more so today, if only due to the canonical formulation inserting the potentials into Schr\"odinger's equation. Nevertheless, also in the Feynman lectures, there is no mention of the problem of gauge-underdetermination that arises from taking the potentials as physical, though different gauge choices are being used. Hence it is not at all clear what is meant be `real' here.

Let us briefly survey Healey's positive account  of the AB effect, which is called the holonomies view \cite{HealeyBook} (also see Belot \cite{Belot} and Wu and Yang \cite{WuYang}). In addition to assigning vectors such as $\textbf{E}$ and $\textbf{B}$ to points in space, this view assigns complex numbers---called \textit{holonomies}---to closed curves in space. These holonomies  are gauge-invariant complex numbers of unit modulus: 
\begin{equation}\label{holo}
h(\gamma)= \exp\blockbrack{i \oint_\gamma \textbf{A} \cdot d\textbf{x}}.
\end{equation}
The holonomies \eqref{holo} are gauge-invariant for the same reason that Eq. \eqref{vanish} is gauge-invariant, using the familiar but tailor-made gauge symmetry of the Narrow Equivalence Class. The loop integrals of the potentials are promoted to fundamental ontology \cite[section 4.4]{HealeyBook}. There are several costs to this approach. Notably, it thus seems one cannot  write down the equations of motion in terms of the holonomies. Also, even though the holonomies act locally, they lead to a non-local theory in the sense that it is non-separable: determining the state in a given region involves the value of $h(\gamma)$ for every possible loop. It is non-separable in the sense that the state of a region is not fully determined by the conjunction of states on all its subregions (which is the kind of non-locality or `non-separability' that Einstein objected to \cite{Einstein}: cf. section \ref{Locality}). 

Healey dismisses the potentials view by claiming that ``there is reason to doubt that the magnetic vector potential is a physically real field," since ``$\textbf{A}$ is not gauge-invariant" \cite[p.22]{Healey}. This kind of gauge-invariance is Healey's \textit{desideratum}. But from this it does \textit{not} follow that one should doubt the physicality of $\textbf{A}$. He argues that, since one can transform away part of the vector potential in the  region outside of the solenoid, for example gauging it to zero at path $1$ of the electron beam (cf, footnote \ref{Binney}), the effect should not derive from $\textbf{A}$ itself. Healey specifically suggests that
\begin{quote} the potential is defined only up to a gauge-transformation, and for any continuous path from source to screen that does not enclose the solenoid there is a gauge-transformation that equates the value of $\textbf{A}$ at every point on that path when a current is flowing to its value when no current is flowing. The shift interference pattern cannot therefore be produced by a direct interaction between individual electrons following such continuous paths and the magnetic vector potential $\textbf{A}$ outside the solenoid. Accepting the physical reality of the vector potential fails to render the AB effect local: while denying its physical reality leaves one without any local explanation of the effect \cite[p.22]{Healey}.
\end{quote}
 Here, Healey seems to be dismissing the potentials view as a sleight of hand---denying it from the start. Healey seems to be saying that there is always a gauge choice in which the explanation cannot be given in terms of a local effect of the vector potential.   Healey's point of view, then, is that if $\phi$ and $\textbf{A}$ are real then they should be real in \textit{all} gauges. But this prompts the question why we use the gauge symmetries that we use.  
  
 One need not follow Healey in ignoring the possibility of additional physical degrees of freedom of the Narrow  Equivalence Class. It is unproblematic to say that only gauge-invariant quantities can be considered physical, but this depends on which definition of the gauge equivalence class one takes, for example the Wide Equivalence Class ($\textbf{A}\mapsto\textbf{A}':=\textbf{A}+\textbf{C}$ such that $\nabla \times \textbf{C}=0$) or the Narrow Equivalence Class ($\textbf{A}\mapsto\textbf{A}':=\textbf{A}+\nabla \chi$), or even a narrower equivalence class ($\textbf{A}\mapsto\textbf{A}':=\textbf{A}+\nabla \chi$ plus a further constraint). In addition to that, on the potentials view, there is in fact a local explanation of the AB phase shift using a narrower equivalence class. I discuss this in sections \ref{Locality} and \ref{Classes}, and in section \ref{Calc} I argue for a specific narrower `Lorenz Equivalence class', where the AB phase shift can be explained ``by a direct interaction between individual electrons following such continuous paths and the magnetic vector potential $\textbf{A}$ outside the solenoid."

A final topic that, I suggest, could be emphasized more strongly in the literature is how the potentials view combines with quantum mechanics. Indeed, there are important constraints on the particular quantum theory one can adhere to, if one simultaneously adopts the potentials view (depending on the version of the potentials view that one is considering). 
  
The choice between quantum theories turns, in part, on whether to include a preferred foliation in a particular quantum theory. That is, if the collapse of the wavefunction is a physically real process, then presumably it picks out some preferred foliation of the spacetime, namely the hyperplane in which the state vector projects.\footnote{One should, however, be careful before equating the apparent non-locality that derives from violations of the Bell inequalities with the apparent non-locality that we have seen in the AB effect above. One can debate whether to begin the inquiry with analogies to violations of the Bell inequalities or the interpretation of gauge freedom. The former position is defended by Healey \cite{Healey} \cite{Healey2} and the latter by Maudlin \cite{Maudlin}.} In such a case, one might pursue the strategy of taking a non-local gauge potential theory as real and aligning the non-local foliation of that gauge theory to the foliation suggested by one's quantum theory. This would, I believe, be a good strategy for reconciling particular collapse theories such as GRW theory with the potentials view. One can also consider the Coulomb gauge,  which both picks out a relativistic structure, as seen in Eq. \eqref{MaxAmpPotCoulomb}, \textit{and} a  preferred foliation, as we have recognized through Eq. \eqref{GaussPotCoulomb}. Following Maudlin's suggestion, this spacetime structure serves well to implement Bohmian mechanics \cite[p.20]{Maudlin2}.

 \section{\textit{Desiderata}: Bell locality and signal locality} \label{Locality}
One reason why gauge-underdetermination is so often only briefly discussed, is that it is not straightforward which gauge should be preferred. As with every occurrence of underdetermination one needs additional principles if one wants to cut down on possibilities. But on what criterion should the gauge be fixed? \textit{I propose that, faced with the original motivation of formulating a local explanation of the AB effect, the choice should be made on the same grounds.}  

Yet clearly locality is itself a vague concept, with several uses, and these uses in turn tend to be vague. Einstein was concerned with the localization of physical facts: what we would now call separability or anti-holism \cite{Einstein}. This still allows for instantaneous action at a distance. The denial of action at a distance gives one the stronger notion of signal locality. I deem signal locality to be successful for our purposes and will discuss it below. First, however, I discuss the more precise notion of a \textit{Bell local theory}, which seems to be the most promising version. Nevertheless, I will come to the conclusion that Bell locality is generally not helpful. Only for some complete gauge choices can we evaluate it.

John Bell regarded gauge choices, as (\textbf{E},\textbf{B})-theorists do, as conventions. His famous example \cite[p.219]{BellLNC} is that of British sovereignty: ``when the Queen of England dies in London, the Prince of Wales becomes \textit{instantaneously} King." But nobody is particularly bothered by this kind of non-locality, since the property of being the British sovereign is not a local beable, nor any kind of beable: it is a convention. In this Pickwickian sense, conventions can indeed travel faster than light. In his 1976 paper on local beables, Bell writes
  \begin{quote} In Maxwell's electromagnetic theory,  for  example,  the fields E and [B] are `physical' (beables, we will say) but the potentials A and $\phi$ are `non-physical'.  Because of gauge invariance the same physical situation can be described by very different potentials. It does not matter that in Coulomb gauge the scalar potential propagates with infinite velocity [see Eq.~\eqref{GaussPotCoulomb}]. It is not really supposed to \textit{be} there. It is just a mathematical convenience \cite[p.54]{BellTLB}."
  \end{quote} 
On the potentials view, however, $\phi$ and $\textbf{A}$ \textit{are} supposed to really be there. In that case, it is not so straightforward to evaluate whether the ($\phi$,\textbf{A})-theory is local in the sense Bell was concerned with---to be explicated below. Different gauge choices lead to different locality considerations, so that gauge choices lead to conceptually different theories. Only in the context of a single clearly defined theory can we evaluate Bell locality. When there is gauge-underdetermination, the choice of gauge (partially) specifies what the supposed ontology (i.e., Bell's local beables) of the theory are. 
 
\begin{figure}
\centering
\includegraphics[width=0.8\textwidth]{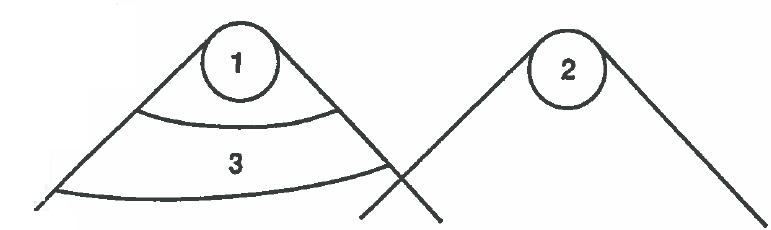}
\caption{\footnotesize{``\textit{Full specification of what happens in 3 makes events in 2 irrelevant for predictions about 1 in a locally causal theory.}" Figure from \cite[p.225]{BellLNC}.}} \label{fig:2}
\end{figure}

So let us define the concept of Bell locality. Consider two spatially separated regions, region $1$ and region $2$, and the events in their respective past light cones. Then consider a third region, region $3$, confined to the past light cone of region $1$, but \textit{completely excluded from the past light cone of region $2$}, as depicted in Fig.~\ref{fig:2}. The idea is that a theory is `Bell local' if and only if events in $2$ are irrelevant for predictions about events in $1$ given that one has a full specification of what happens in region $3$. Mathematically, the conditional probability that an event $\mathcal{E}_1$ in region $1$ occurs given the set of events $\mathcal{C}_3$ in region $3$ should be equal to the conditional probability that the same event $\mathcal{E}_1$ occurs given the same $\mathcal{C}_3$ and also any additional event $\mathcal{E}_2$ in region $2$: 
\begin{equation} \label{local}
P[\mathcal{E}_1 | \mathcal{C}_3] = P[\mathcal{E}_1 | \mathcal{C}_{3},\mathcal{E}_2].
\end{equation}
Of course, $\mathcal{E}_1$ and $\mathcal{E}_2$ might still be correlated events. After all, their respective past light cones do have an overlapping region which can contain  an event that is a common causal influence on the two events. There can also be some indirect influence of the causes of $\mathcal{E}_1$ overlapping with the causes of $\mathcal{E}_2$. But the point is that \textit{in a Bell local theory} any such information that event $\mathcal{E}_2$ might reveal about event $\mathcal{E}_1$ is already contained in $\mathcal{C}_3$: so given $\mathcal{C}_3$, any information about whether $\mathcal{E}_2$ occurs is redundant.\footnote{To simplify terminology, I use the property `Bell local' as synonymous with `locally causal', interpreting Eq. \eqref{local} as $\mathcal{C}_3$  `completely screens off $\mathcal{E}_1$ from events in $\mathcal{E}_2$'.}\\
\indent Note further that Bell locality is a property at the level of the theory and does not directly mention events  in reality. To link the theory to the things in the world, Bell introduces \cite[p.219]{BellLNC} the notion of \textit{beables}: the ``\textit{be}ables of the theory are those entities in it which are, at least tentatively, to be taken seriously, as corresponding to something real."  Hence, beables are theoretical entities that can be directly linked to something in the world, depending on how one fleshes this out (cf. footnote~\ref{link}). Further, beables are \textit{local} beables if they can be confined to some finite spacetime region. One can see that Bell locality presupposes separability.

To further examine the usefulness of Bell locality in the context of the AB effect, one should realize that quantum mechanics itself is Bell \textit{non}-local, due to the violations of the Bell inequalities. As mentioned earlier in this section, one can compare the relation between different senses of non-locality in the AB experiment and the experimental tests of the Bell inequalities. How severe Bell non-locality really is remains widely contested, especially concerning whether parameter independence or outcome independence is violated, which depends on making Bell's work even more precise   \cite{Seevinck}. Furthermore, to evaluate the Bell locality of a theory, we presuppose a measurement-problem-free theory where the ontological commitments are reasonably clear. That is, we cannot see  at the phenomenological level if outcome independence or parameter independence is violated, since we inevitably run into the measurement problem when we consider `outcomes' \cite{Jeremy}. Yet, this leads us astray in our present context. In this present context the goal is to formulate \textit{desiderata} in terms of locality considerations. That is, we attempt to \textit{use locality as a criterion for theory choice in a gauge-underdetermined situation so as narrow down the class of admitted gauge degrees of freedom}.   In any case, I maintain that matters are worsened if there is an additional beable in the theory that violates Eq.~\eqref{local}. For example, a Coulomb gauge potentials theory would violate it through  direct action at a distance (see  below).

The electric and magnetic fields are defined in four-dimensional spacetime and satisfy the wave equation, Eq.~\eqref{Ewave}, which automatically confines the propagation of causal influences to the light cones. Since the theory is deterministic, $\mathcal{C}_3$ uniquely determines event $\mathcal{E}_1$, since it is in the future domain of dependence $\Sigma(\mathcal{C}_3)$ of region $\mathcal{C}_3$: a sufficient set of relevant initial conditions for $\mathcal{E}_1$ lie in region $3$ so that the equations of motion of all the beables involved in $\mathcal{E}_1$ have a unique solution. On the fields view (i.e., taking the electric and magnetic fields as physical), the (\textbf{E},\textbf{B})-theory is Bell local and the \textbf{E} and \textbf{B} fields are local beables.

Yet, even though the (\textbf{E},\textbf{B})-theory is Bell local, it is unable to give a local explanation of the AB phase shift. It would therefore be natural to evaluate the Bell locality of the potentials view in the context of the AB effect. However, the experimental set-up is not easily translated into the language of Eq.~\eqref{local}. In the AB experiment the electron beam comes arbitrarily close to the solenoid, so that two putative events $\mathcal{E}_1$ and $\mathcal{E}_2$ involving the electron cannot be spacelike separated without destroying the AB phase shift. This is because the electrons need to form a closed loop and the electrons are confined to the light cone. In short: the Aharonov-Bohm set-up involves a single system, making Bell locality hard to apply. Therefore, Bell locality will not lend itself as a criterion to  pick out some local gauge choice. One can only take a particular potentials theory, put it into the language of Eq.~\eqref{local}, and consider if there are violations of Bell locality.

 The sense of locality that, I will argue, can tackle gauge-underdetermination and provide a local picture of the AB effect is that of signal locality. This is what one usually understands by the concept `locality': it means that the causal influences that bodies exert on one another propagate at some finite speed. So the theory is to treat causal influences as propagating at a finite speed through (disjoint regions of) spacetime. One imagines `messengers' that take some time to travel from one event to the other in order to `tell' the actors there how they should react. If that travelling speed is infinite, one speaks of `action at a distance'. 

 Electric and magnetic fields satisfy signal locality. Taking the the curl of both Faraday's law and the Maxwell-Amp\`ere law, one derives that the electric and magnetic fields both satisfy the wave equation:
\begin{equation} \label{Ewave}
\nabla^2 \textbf{E} = \frac{1}{c^2}\frac{\partial ^2 \textbf{E} }{\partial t^2},
\end{equation} and likewise for the B-field. The solutions are waves moving at speeds $c$, such that electromagnetic signals travel with a finite velocity.\footnote{This is also the case when sources are included. Then, the wave equations acquire an inhomogeneous part, which can be solved by the Green's function method. A point charge (or current) at position $\textbf{x}'$ considered only at one point in time ($t'=0$) gives rise to an electric (or magnetic) field, evolving forward in time,  $\textbf{E}(\textbf{x},t)= - \delta\brack{t-|\textbf{x}-\textbf{x}'|/c}/4 \pi |\textbf{x}-\textbf{x}'|\hat{r}$, with the delta distribution $\delta(x)$ and radial unit vector $\hat{r}$.  It is interpreted as  a wavefront that propagates spherically outwards, reaching all the positions $\textbf{x}$ at times  $t=\textbf{x}/c$. A real source is represented by the superposition of such charges (currents) in space and time. \label{h}} 

In the Coulomb gauge---as we have seen in Eq.~\eqref{GaussPotCoulomb} and in Bell's words above---the scalar potential satisfies the Laplace equation, which is a non-local equation also satisfied by the Newtonian gravitational potential. One way to see this is to take the limit $c \rightarrow \infty$ in the wave equation so that it reduces to Laplace's equation. Hence, signal locality is violated.
  
 If we now briefly return to the property of Bell locality, we recognize that only now that a complete gauge is fixed, viz. the Coulomb gauge, one is in a position to evaluate this in terms of Bell locality. It is readily seen that Eq.~\eqref{local} is violated since the beables at $\mathcal{E}_1$ are not fully specified by region  $\mathcal{C}_3$ and can be spacelike influenced by $\mathcal{E}_2$ from outside the past light cone. Hence, as the scalar potential is a beable in this theory, the Coulomb gauge potentials theory ($\phi$,$\textbf{A})_{\text{CG}}$ violates Bell locality in quite a crude way;  indeed, exactly as crudely as Newtonian gravitational theory does.
  
\begin{figure}
\centering
\includegraphics[width=1\textwidth]{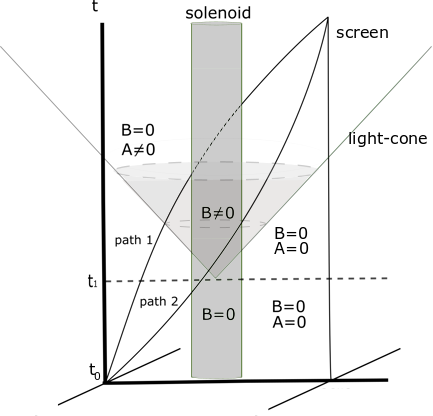}
\caption{\footnotesize{A sketch of the AB effect. The vector potential $\textbf{A}$, in the Lorenz gauge potentials theory ($\phi$,\textbf{A})$_{LG}$, propagates outward at speed $c$ after the current has been switched on (instantaneously) at $t_1$, while the magnetic field $\textbf{B}$ is confined to the region occupied by the solenoid at all times.   An electron wave-packet is emitted at $t_0$ and is steered by external potentials around the solenoid both \textit{into the plane of the figure} or away from the reader and \textit{out of the plane of the figure} or towards the reader. For simplicity, the slits and surface of the screen are not drawn, but all possible paths of the electron must be imagined. If one takes the vector potential in the Lorenz gauge as fundamental ontology, then there is a signal-local explanation of the phase shift (while there is none using the Wide or Narrow Equivalence Class). The resulting gauge equivalence class, here dubbed the Lorenz Equivalence Class, has a residual gauge symmetry in the form of a scalar wave, i.e., $\chi$ satisfying $\nabla^2 \chi =c^{-2}\partial^2 \chi / \partial t^2$. This narrows down the usual equivalence class  $\textbf{A} \mapsto \textbf{A}':=\textbf{A}+\nabla \chi$ significantly.  \label{fig:3}}}
\end{figure}

  Alternatively, in the Lorenz gauge potentials theory ($\phi$,\textbf{A})$_{LG}$, both the scalar potential $\phi$ and the vector potential $\textbf{A}$ satisfy the wave equation and propagate with velocity $c$, as exhibited by Eqs. \eqref{GaussPotLorenz}-\eqref{MaxAmpPotLorenz}. Hence, in the Lorenz gauge, signal locality is satisfied, as can be motivated by Fig.~\ref{fig:3}. So here is a local explanation that---unlike the (\textbf{E},\textbf{B})-theory---has the ability to explain the AB phase shift locally. The next two concluding sections develop this theme.

\section{Local interaction and narrowing equivalence classes} \label{Classes}
Ahanorov and Bohm speak of the apparent non-locality of their experiment as follows.
\begin{quote}
Of course, our discussion does not bring into question the gauge invariance of the theory. But it does show that in a theory involving only local interactions (e.g., Schrodinger's or Dirac's equation, and current quantum-mechanical field theories), the potentials must, in certain cases, be considered as physically effective, even when there are no fields acting on the charged particles \cite[p.490]{AB}.
\end{quote}
Hence they consider the results inexplicable in terms of local interactions, as long as the potentials remain interpreted as mathematical fictions. For the behaviour of the quantum probe would be influenced by the presence of $\textbf{B}$ inside the solenoid, although $\textbf{B}$ vanishes at every point in space outside the solenoid at every  time that the probe is present at those spatial points. 

Aharonov and Bohm  argue for locality by noting that ``according to our current relativistic notions, all fields must interact only locally \cite[p.490]{AB}." As their paper pre-dates Bell's formulation of locality, and they do not explicate what they mean by locality, it is not clear if Aharonov and Bohm have something like local beables in mind. It is not even clear if they have signal locality in mind, although they can certainly be read that way. At the very least, their notion of locality seems more in line with the idea that \textit{a theory is local when given in terms of equations of motion where the interactions only take place when the coordinates of the interacting systems coincide}. Call this \textit{local interactions}. 

Coming to what was promised at the end of section \ref{GU}, one can  read Aharonov and Bohm conservatively: as rejecting the Wide Equivalence Class that leaves only the $\textbf{E}$ and $\textbf{B}$ fields invariant, namely Eqs. \eqref{gaugevector}-\eqref{gaugevector2}, repeated here as $\textbf{A} \mapsto \textbf{A}':=\textbf{A}+\textbf{C}$ and $\phi \mapsto \phi':=\phi-C_0, $ such that $\nabla \times \textbf{C}=0$ and $\nabla C_0 = \partial \textbf{C}/ \partial t$. One can now reformulate Aharonov and Bohm's legacy: the traditional fields view that used the Wide Equivalence Class to encode the physical states of affairs in the world is inadequate because the class is \textit{too} wide. For, in general, two elements of such a class differ in their physics, as is revealed by the shifted interference patterns exhibited by probing the field around the solenoid. The Narrower Equivalence Class joins these elements together. The `puncture' in space that is the solenoid, introduces a multiply-connected space and the difference brought about by this non-trivial topology can be probed by a suitable quantum particle.\footnote{The electron in the AB experiment is described by a quantum wave function. In some sense, however, this is only accidental. If one had used a classical, not a quantum, probe, an effect by the potential on the probe would still be there, but there might not be such a straightforward way to observe that effect. In addition---even though this does not demonstrate that the quantity is classical---note that computing the AB phase shift does not involve any loop integrals. See also  Chapter 8 of Neil Dewar's forthcoming book \cite{Dewar} and Henrique Gomes \cite[p.19]{Henrique}.} In other words, on a non-trivial topology the Narrow Equivalence Class is not the same as an electric and magnetic fields.  The former has additional degrees of freedom, and these are the ones that account for the AB effect in terms of local interactions of the fields.\footnote{Much more can be said about the specific considerations of Bohm and Aharonov (cf.~ the `Vaidman-ACR'-debate in footnote~\ref{VaidmanACR} and references there).} 

As I have stressed, there are different senses of locality. Even someone seeking a local explanation in a strong sense must admit that a space such as the Moebius strip can have global features that are not caught by the description of any local patch. Hence, locally you cannot say if the space is simply-connected or multiply-connected. On the other hand, everybody must accept that the value of a line integral such as $\oint_{\gamma}\textbf{A} \cdot d\textbf{x}$ changes incrementally as the endpoint varies. So, for the AB effect, anybody seeking a local explanation must respect such constraints. Yet, one might wonder what happened to the signal locality that we remember from the fields view, namely that $\textbf{E}$ and $\textbf{B}$ propagate as waves with speed $c$.   Are we forced to give this up or can we have it on top of the above notion of local interaction? In other words, what does a signal-local explanation of the AB effect look like?

\section{A suggestion: the Lorenz Equivalence Class} \label{Calc}
In the final paragraph of section \ref{Locality}, I pointed out that the Lorenz gauge potentials theory ($\phi$,\textbf{A})$_{LG}$ satisfies signal locality. It does so because in the Lorenz gauge the equations of motion of $\phi$ and $\textbf{A}$ satisfy the wave equation \eqref{GaussPotLorenz}-\eqref{MaxAmpPotLorenz} with propagation speed $c$. Hence, if we narrow the equivalence class further by using the Lorenz gauge condition, the potentials can be interpreted as travelling waves, analogous to the travelling of $\textbf{E}$ and $\textbf{B}$. Then, some degrees of freedom of the potentials---in addition to those giving rise to $\textbf{E}$ and $\textbf{B}$---travel outward from the solenoid to the electron. Thus, this ($\phi$,\textbf{A})$_{LG}$-theory opens the door to a signal-local explanation of the AB phase shift, as illustrated  in Fig.~\ref{fig:3}.

My proposal is, accordingly, to take the even narrower gauge equivalence class than the Narrow Equivalence Class by adding the constraint on the gauge function $\chi(\textbf{x},t)$ in the form of the Lorenz gauge condition \eqref{LorenzGauge}.  This constraint of the last line can be readily  calculated by performing a gauge transformation \eqref{gauges}-\eqref{gauges2} of the condition \eqref{LorenzGauge}. It turns out that there is a residual freedom in which $\chi$ itself obeys the wave equation.\footnote{Alternatively, using the gauge freedom of the Wide Equivalence Class \eqref{gaugevector}-\eqref{gaugevector2} returns four wave equations, $\nabla^2 C_0= (1/c^2) \partial^2 C_0 /  \partial t^2 $ and $\nabla^2 \textbf{C}= (1/c^2) \partial^2 \textbf{C}  /  \partial t^2 $, still subject to $ \partial \textbf{C} / \partial t = \nabla C_0$. However, this Wide Equivalence Class + Lorenz gauge condition is not sufficient to make the line integral \eqref{vanish} independent of the path, which is what I have in section \ref{Classes} called Aharonov and Bohm's legacy.} The \textit{Lorenz Equivalence Class} is then
\begin{align}
&\phi \mapsto \phi':=\phi-\frac{\partial \chi}{\partial t}, \nonumber \\
& \textbf{A} \mapsto \textbf{A}':= \textbf{A}+\nabla \chi, \\
&\nabla^2 \chi=\frac{1}{c^2}\frac{\partial^2 \chi }{ \partial t^2 }.\nonumber
 \end{align}

On this view, the `true' gauge degrees of freedom are the ones that leave the scalar wave solution intact. This scalar wave solution is a result of the Lorenz gauge-fix being incomplete or partial. The remaining freedom describes how the potentials behave far away, i.e. towards spatial infinity. This can be fixed by initial conditions; but, for all I have argued here, those initial conditions seem as if they cannot be determined empirically.
 
Let us return to the spectrum in Figure \ref{fig:spectrum}. We can now recognize that the Lorenz Equivalence Class is quite far to the left within the potentials view, but there is still the residual gauge freedom resulting from the partial gauge-fix, which we can safely regard as mathematical fiction. The Coulomb gauge condition \eqref{CoulombGauge} is a complete gauge-fix, as it would result in the Laplace equation for the gauge function, $\nabla^2 \chi =0$, which---for given boundary conditions that one can take to be given by the experiment---has a unique solution. Another oft-used way to achieve this is the Lorenz gauge plus the `temporal gauge' $\phi=0$. Were one to gauge-fix completely along such routes, the resulting philosophical position is the One True Gauge principle, which can be seen as a gauge equivalence class with one member. 

To be precise, however, one leaves room for the trivial gauge equivalence class, which consists of transformations of the unit one expresses the potentials in, call it the `Singleton Set'. Although this issue is hardly ever discussed in connection with the AB effect and gauge theories---and rightly so, since this variety is conceptually straightforward---nevertheless it means that \textit{until} one chooses units, even a complete gauge-fixing yields equivalence classes each of which has many elements, corresponding to the various possible units that one might choose to adopt. In the fields view, there is the same freedom to change the units of $\textbf{E}$ and $\textbf{B}$, indicating that also the electric and magnetic field are not free from human choice.

\section{Conclusion}
Aharonov and Bohm showed us that there is good reason to take the electromagnetic potentials seriously as part of the fundamental ontology of electrodynamics. One consequence of their work was that the Narrow Equivalence Class has become standard in contemporary textbooks, leading to some physical degrees of freedom in the potentials over and above those that give rise to $\textbf{E}$ and $\textbf{B}$. The reality of the gauge potential demands careful treatment of one's commitment to the gauge symmetry of the theory. Naively taking the potentials as `real' immediately runs into an indeterminism of the equations of motion of the potentials and hence a group of rival theories between which one cannot choose on empirical grounds: gauge-underdetermination. This is a partial commitment: one cannot have the reality of the potential (having the cake) while also regarding gauge-fixing as arbitrary (eating it too). Such a move does not only play a role in  electrodynamics, but is seen in all theories where degrees of freedom can be seen as `gauge', such as all of the Standard Model Lagrangian, diffeomorphism invariance in relativity theory, and various approaches to quantum gravity. 

 One way to have signal locality is to choose the Lorenz gauge: $\phi$ and $\textbf{A}$ then obey the wave equation, just like $\textbf{E}$ and $\textbf{B}$ do. Agreed, besides the Lorenz gauge there are many conceivable positions to lying on the spectrum of Figure \ref{fig:spectrum}, and I suggest these should be explored on their philosophical merits and shortcomings. I have chosen to stay close to the original motivation for taking the potentials as physical, namely locality. Emphasizing the \textit{desideratum} of signal locality, I have suggested we should use the Lorenz gauge, because then the wave equation is satisfied by the potentials. It is true that the dynamics of the potentials in the Lorenz gauge is automatically Lorentz invariant, but this is also the case for other (relativistic) gauges. In conclusion, I thus suggest we should narrow down further the Narrow Equivalence Class to the Lorenz Equivalence Class, with a scalar wave solution that encodes the residual gauge freedom, as the ontology of electrodynamics.
%

\section*{Acknowledgments}
I wish to thank Jeremy Butterfield, Guido Bacciagaluppi and Henrique Gomes for thorough comments and corrections; and Dennis Dieks, Maaneli Derakhshani, Sam Rijken, Vipin Chaudhary and Caspar Jacobs for helpful correspondence and discussions; and also two anonymous reviewers for crucial observations. 

 \footnotesize{ } 

\begin{thebibliography}{1}
 
\bibitem{Dauber} Dauber, J., M. Oellers, F. Venn, A. Epping, K. Watanabe, T. Taniguchi, F. Hassler, and C. Stampfer (2017). ``Aharonov-Bohm oscillations and magnetic focusing in ballistic graphene rings." \textit{Physical Review B} \textbf{96}, 205407.

\bibitem{Batelaan} Batelaan, H. $\&$ A. Tonomura (2009). ``The Aharonov–Bohm effects: Variations on a subtle theme." \textit{Physics Today} \textbf{62}, 9.

\bibitem{Tran} Tran, M. (2018). ``Evidence for Maxwell's equations, fields, force laws and alternative theories of classical electrodynamics." \textit{European Journal of Physics} \textbf{39}, 6.

\bibitem{Kasunic} Kasunic, K. (2019). ``Magnetic Aharonov-Bohm effects and the quantum phase shift: A heuristic interpretation." \textit{American Journal of Physics} \textbf{87}, 745.

\bibitem{Berry} Berry, M. (2010). ``Aptly named Aharonov-Bohm effect has classical analogue, long history." \textit{Physics Today} \textbf{63}, 8.
 
 \bibitem{HealeyBook} Healey, R. (2007). \textit{Gauging What's Real: The Conceptual Foundations of Contemporary Gauge Theories}. Oxford: Oxford University Press. 

\bibitem{Myrvold} Myrvold, W. C. (2011). ``Nonseparability, Classical, and Quantum."\textit{British Journal for the Philosophy of Science} \textbf{62} (2), 417.

\bibitem{Mattingly1} Mattingly, J. (2005). ``Which gauge matters?" \textit{Studies in the History and Philosophy of Modern Physics} \textbf{37}, 243.

\bibitem{Mattingly2} Mattingly, J. (2007). ``Classical fields and quantum time-evolution in the Aharonov–Bohm effect". \textit{Studies in the History and Philosophy of Modern Physics} \textbf{38}, 888.

\bibitem{Boyer2} Boyer, T. (2006). ``Darwin-Lagrangian analysis 
for the interaction of a point charge and a magnet: Considerations related to the controversy regarding the Aharonov-Bohm and Aharonov-Casher phase 
shifts." \textit{Journal of Phys. A: Math. Gen.} \textbf{39}, 3455.

\bibitem{Vaidman2012} Vaidman, L. (2012). ``Role of Potentials in the Aharonov-Bohm Effect." \textit{Physical Review A} \textbf{86}, 040101(R). 

\bibitem{Vaidman2015} Vaidman, L. (2015). ``Reply to a Comment on ``Role of Potentials in the Aharonov-Bohm Effect"." \textit{Physical Review A} \textbf{92}, 026102.

\bibitem{ACR2015} Aharonov, Y., E. Cohen, D. Rohrlich (2015). ``Comment on ``Role of Potentials in the Aharonov-Bohm Effect"." \textit{Physical Review A} \textbf{92}, 026101. 

\bibitem{ACR2016} Aharonov, Y., E. Cohen, D. Rohrlich (2016). ``Nonlocality of the Aharonov-Bohm effect." \textit{Physical Review A} \textbf{93}, 042110. 

\bibitem{Rizzi} Pearle, P. $\&$ A. Rizzi (2017). ``Quantum-mechanical inclusion of the source in the Aharonov-Bohm effects." \textit{Physical Review A} \textbf{95}, 052124.



\bibitem{AB}  Aharonov, Y. $\&$ D. Bohm (1959). ``Significance of Electromagnetic Potentials in the Quantum Theory." \textit{The Physical Review} \textbf{115} (3), 485. 

\bibitem{ES}  Ehrenberg, W. $\&$ R. E. Siday (1949). ``The Refractive Index in Electron Optics and the Principles of Dynamics." \textit{Proceedings of the Physical Society B} \textbf{62} (1), 8.

\bibitem{Peshkin} Peshkin, M. $\&$ A. Tonomura (1989). \textit{The Aharonov-Bohm Effect}.\textit{ Lecture Notes in Physics}, \textbf{340}. Edited by W. Beiglb\"ock, H. Araki, J. Ehlers, K. Hepp, R. Kippenhahn, D. Ruelle, H.A. Weidenm\"uller, J. Wess, Zittartz. Berlin: Spinger-Verlag. 

\bibitem{Dirac} Dirac, P. A. M. (1931). ``Quantised Singularities in the Electromagnetic Field." \textit{Proceedings of the Royal Society} \textbf{A133} (60), 60.

\bibitem{BinneySkinner} Binney, J. $\&$ D. Skinner (2008-2013). \textit{The Physics of Quantum Mechanics}. Cappella Archive.

\bibitem{Ardourel} Ardourel, V. $\&$ A. Guay (2018). ``Why is the transference theory of causation insufficient? The challenge of the Aharonov-Bohm effect." \textit{Studies in History and Philosophy of Science B}, preprint.

\bibitem{Boyer} Boyer, T. H. (1972). ``Misinterpretation of the Aharonov-Bohm Effect." \textit{American Journal of Physics} \textbf{40}, 56.

\bibitem{Quine} Quine, W. O. (1953). ``Chapter 1: On what there is." In: \textit{From a Logical Point of View}. Cambridge, MA: Harvard University Press. 

\bibitem{Allori} Allori, V. (2015). ``Primitive Ontology in a Nutshell." \textit{International Journal of Quantum Foundations} \textbf{1} (3), 107. 

\bibitem{Ladyman} Ladyman, J. (1998). ``What is structural realism?" \textit{Studies in History and Philosophy of Modern Science} \textbf{29}, 409.

\bibitem{Bladel} Bladel, van, J. V. (1991). ``Lorenz or Lorentz?" \textit{IEEE Antennas and Propagation Magazine} \textbf{32} (2), 69.

\bibitem{Norton}  Norton, J. D. (2015). ``The Hole Argument." \textit{The Stanford Encyclopedia of Philosophy}. Edited by E. N. Zalta, section 10.3.2. 

\bibitem{Maudlin} Maudlin, T. (1998). ``Healey on the Aharonov-Bohm effect."  \textit{Philosophy of Science} \textbf{65}, 361.

\bibitem{Norsen}  Norsen, T. (2017). \textit{Foundations of quantum mechanics. An exploration of the physical meaning of quantum theory}. Springer International Publishing.

\bibitem{Moriyasu} Moriyasu, K. (1983). \textit{An elementary primer for gauge theory}. Singapore: World Scientific Publishing.

\bibitem{FeynmanLec} Feynman, R. P., R. B. Leighton, M. Sands (1965). \textit{Feynman Lectures on Physics Volume 2}. Reading, MA: Addison-Wesley.

 \bibitem{Belot} Belot, G. (1998). ``Understanding Electromagnetism." \textit{British Journal for Philosophy of Science} \textbf{49}, 531.

\bibitem{WuYang} Wu, T. T. $\&$ C. N. Yang (1975). ``Concept of Nonintegrable Phase Factors and Global Formulation of Gauge Fields." \textit{Physical Review} D \textbf{12}, 3845--3857.

\bibitem{Einstein} Einstein, A. (1948). ``Quantum Mechanics and Reality." \textit{Dialectica} \textbf{2}, 320.

\bibitem{Healey} Healey, R. (1997). ``Nonlocality and the Aharonov-Bohm Effect." \textit{Philosophy of Science} \textbf{64} (1), 18.

\bibitem{Healey2} Healey, R. (1999). ``Quantum Analogies: A Reply to Maudlin."  \textit{Philosophy of Science} \textbf{66} (3), 440.

\bibitem{Maudlin2} Maudlin, T. (2018). ``Ontological Clarity via Canonical Presentation: Electromagnetism and the Aharonov-Bohm Effect." \textit{Entropy}  \textbf{20}, 465.

\bibitem{BellLNC} Bell, J. S. (1990). ``La Nouvelle Cuisine." In: \textit{John S. Bell on the Foundations of Quantum Mechanics}, 216--234. Edited by K. Gottfried $\&$ M. Veltman. Singapore: World Scientific Publishing.

\bibitem{BellTLB} Bell, J. S. (1976). ``The theory of local beables."  \textit{Epistemological Letters} \textbf{9}, 11.

\bibitem{Seevinck}  Seevinck, M. P. (2010). ``Not throwing out the baby with the bathwater: Bell's condition of local causality mathematically `sharp and clean'." In: \textit{Explanation, Prediction, and Confirmation}, 25. Edited by D. Dieks, S. Hartmann, T. Uebel, M. Weber, W. J. Gonzalez.

\bibitem{Jeremy}  Butterfield, J. N. (2018). ``Peaceful Coexistence: Examining Kent's Relativistic Solution to the Quantum Measurement Problem." \textit{Springer Proceedings in Mathematics and Statistics} \textbf{261}, 277.

\bibitem{Dewar} Dewar, N. (Forthcoming). \textit{Structure and Equivalence}. Accessible at LSE/Cambridge Philosophy of Physics Bootcamp: \url{https://personal.lse.ac.uk/robert49/PPB/}.

\bibitem{Henrique} Gomes, H. (2019). ``Holism as the significance of gauge symmetries." 
 	\textit{ArXiv}: 1910.05330 [physics.hist-ph]: \url{https://arxiv.org/abs/1910.05330}.

\bibitem{Dewar2} Dewar, N. (2019). ``Sophistication about Symmetries." \textit{British Journal Philosophy of Science } \textbf{70} (2), 485.
  \end{thebibliography}
\end{document}